\documentclass[acmsmall]{acmart}

\usepackage[utf8]{inputenc} 
\usepackage[T1]{fontenc}    
\usepackage{hyperref}       
\usepackage{url}            
\usepackage{booktabs}       
\usepackage{amsfonts}       
\usepackage{nicefrac}       
\usepackage{microtype}      
\usepackage{amsmath,amsfonts}
\usepackage{algorithmic}
\usepackage{graphicx}
\usepackage{textcomp}
\usepackage{xcolor}
\usepackage{amsthm}
\usepackage{enumerate}
\usepackage{multirow}
\usepackage{caption}
\usepackage{subfigure}
\usepackage{array}
\usepackage{tabularx}
\usepackage{tabu}
\usepackage{wrapfig}

\usepackage{algorithm}

\AtBeginDocument{%
  \providecommand\BibTeX{{%
    \normalfont B\kern-0.5em{\scshape i\kern-0.25em b}\kern-0.8em\TeX}}}

\acmJournal{TDS}

\begin{document}

\title{Adversarial Energy Disaggregation for Non-intrusive Load Monitoring}

\author{Zhekai Du}
\affiliation{%
  \institution{School of Computer Science and Engineering, University of Electronic Science and Technology of China}
  \streetaddress{2006 Xiyuan Ave., High-tech Zone West}
  \city{Chengdu}
  \state{Sichuan}
  \country{China}
  \postcode{611731}
}

\author{Jingjing Li}
\authornotemark[1]
\email{Corresponding author, lijin117@yeah.net}
\affiliation{%
  \institution{School of Computer Science and Engineering, University of Electronic Science and Technology of China}
  \streetaddress{2006 Xiyuan Ave., High-tech Zone West}
  \city{Chengdu}
  \state{Sichuan}
  \country{China}
  \postcode{611731}
}

\author{Lei Zhu}
\affiliation{%
  \institution{School of Information Science and Engineering, Shandong Normal University}
  \streetaddress{No.1, University Road, Science Park, Changqing}
  \city{Jinan}
  \state{Shandong}
  \country{China}
  \postcode{250358}
}

\author{Ke Lu}
\affiliation{%
  \institution{School of Computer Science and Engineering, University of Electronic Science and Technology of China}
  \streetaddress{2006 Xiyuan Ave., High-tech Zone West}
  \city{Chengdu}
  \state{Sichuan}
  \country{China}
  \postcode{611731}
}

\author{Heng Tao Shen}
\affiliation{%
  \institution{School of Computer Science and Engineering, University of Electronic Science and Technology of China}
  \streetaddress{2006 Xiyuan Ave., High-tech Zone West}
  \city{Chengdu}
  \state{Sichuan}
  \country{China}
  \postcode{611731}
}

\renewcommand{\shortauthors}{Zhekai Du et al.}

\begin{abstract}
  Energy disaggregation, also known as non-intrusive load monitoring (NILM), challenges the problem of separating the whole-home electricity usage into appliance-specific individual consumptions, which is a typical application of data analysis. {NILM aims to help households understand how the energy is used and consequently tell them how to effectively manage the energy, thus allowing energy efficiency which is considered as one of the twin pillars of sustainable energy policy (i.e., energy efficiency and renewable energy).} Although NILM is unidentifiable, it is widely believed that the NILM problem can be addressed by data science. Most of the existing approaches address the energy disaggregation problem by conventional techniques such as sparse coding, non-negative matrix factorization, and hidden Markov model. Recent advances reveal that deep neural networks (DNNs) can get favorable performance for NILM since DNNs can inherently learn the discriminative signatures of the different appliances. In this paper, we propose a novel method named adversarial energy disaggregation (AED) based on DNNs. We introduce the idea of adversarial learning into NILM, which is new for the energy disaggregation task. Our method trains a generator and multiple discriminators via an adversarial fashion. The proposed method not only learns shard representations for different appliances, but captures the specific multimode structures of each appliance. Extensive experiments on real-world datasets verify that our method can achieve new state-of-the-art performance.
\end{abstract}

\ccsdesc[300]{Applied computing~Industry and manufacturing}
\ccsdesc[500]{Information systems~Data mining}
\ccsdesc[500]{Information systems~Data streaming}

\keywords{energy disaggregation, non-intrusive load monitoring (NILM), signal processing, data analysis}

\maketitle
 
\section{Introduction}
\begin{figure*}[t]
\begin{center}
\includegraphics[width=1\linewidth]{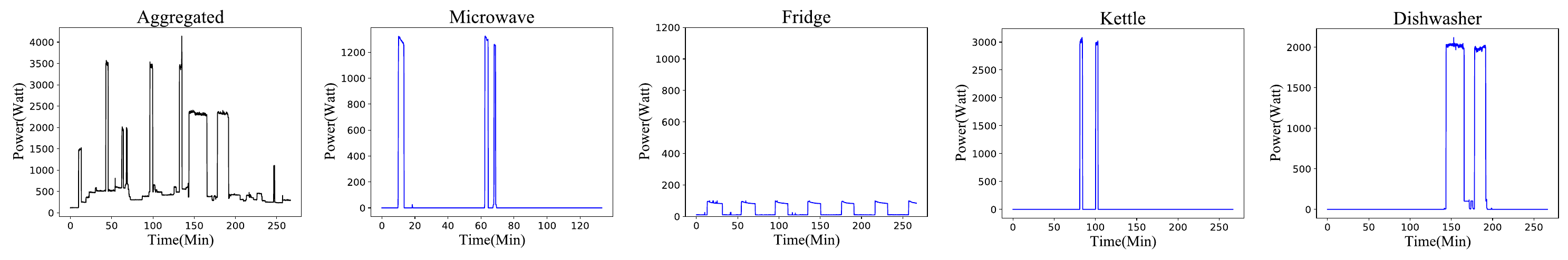}
\end{center}
\caption{An illustration of the task in this paper. The sub-figures (from left to right) show the aggregated signal, the signals of microwave, fridge, kettle and dishwasher, respectively. The task of energy disaggregation is to decompose the aggregated signal into different appliance-specific signals.}
\label{fig:task}
\end{figure*}

{E}{nergy} efficiency has been recognized as one of the twin pillars of sustainable energy~\cite{prindle2007twin}, which is a great challenge facing humanity in the 21st century~\cite{rosenzweig2008attributing}. Nowadays, most of the energy consumption behaviors have digital records and many energy problems can be formulated as informatics problems~\cite{kolter2010energy}. Thus, there is a growing expectation in our community that if data science can play a role in addressing the energy challenge. In this paper, we focus on energy disaggregation, a.k.a. non-intrusive load monitoring (NILM)~\cite{hart1992nonintrusive}, which is proven to have significant effects on energy efficiency~\cite{darby2006effectiveness,neenan2009residential}. Specifically, energy disaggregation investigates the problem of separating an aggregated energy signal into the individual consumptions of different appliances. For a better understanding, we give an example of the NILM task in Fig.~\ref{fig:task}. Previous studies~\cite{darby2006effectiveness,neenan2009residential,fischer2008feedback} reported~that estimating the appliance-level energy consumption data is able to reduce the energy consumption by as~much~as~15\%.

Technically, energy disaggregation can be formulated as a single-channel blind source separation (BSS) problem~\cite{wang2009nonnegative,lee2000ica}. This problem is not trivial since it is unidentifiable, {i.e., one needs to discover more than one source from a single observation. It was originally introduced by Hart in 1990s \cite{hart1992nonintrusive} and has been an active research topic since then. As the demand of energy management increasing continuously, fruitful line of works have studied to address the NILM problem over the decades. A majority of existing methods either leverage the idea of signal processing that explicitly resorts to the features of appliances \cite{wang2018non, berges2011user, du2019multi, jin2011time}, or using machine learning methods in supervised \cite{kelly2015neural, zhang2018sequence, shin2019subtask} and unsupervised manners \cite{zhong2014signal, kolter2012approximate}. Specifically, many machine learning methods model the energy consumption of appliances by unsupervised probabilistic approaches such as factorial hidden Markov models (FHMM)~\cite{zhong2014signal} and its variants \cite{kolter2012approximate}, other methods also deploy machine learning techniques such as sparse coding~\cite{kolter2010energy}, matrix factorization \cite{batra2017matrix} and $k$-nearest neighbor ($k$-NN)~\cite{tabatabaei2016toward} to separate the energy signal. Most of these methods are event-driving, i.e., they track the footprints of different appliances to estimate whether a device is turn on or off. Recently, with the releases of many large-scale public datasets \cite{kolter2011redd, kelly2015uk}, deep neural networks (DNNs) have been successfully applied to energy disaggregation in a supervised way. These methods are more like model-driving, i.e., they try to automatically reveal the temporal structure embedded in the observation data. For instance, Neural NILM~\cite{kelly2015neural} introduces sequence-to-sequence (seq2seq) learning into energy disaggregation and achieves remarkable performance improvement against traditional methods.}

Inspired by the success of DNNs in energy disaggregation, we propose a novel method based on DNNs in this paper. Specifically, we introduce adversarial learning into energy disaggregation. The idea of adversarial learning was reported in generative adversarial networks (GANs)~\cite{goodfellow2014generative}. GANs jointly train a generator and a discriminator via an adversarial manner. The generator synthesizes samples from noises to fool the discriminator. At the same time, the discriminator distinguishes whether a sample is real or fake. Once the discriminator is confused, it is considered that the generated samples have the same distribution as the real ones. The basic idea of our adversarial energy disaggregation (AED) is similar to GANs. Suppose that there is an expert (discriminator) who can recognize a specific electric appliance by its signal, once the signal features learned by our model (generator) can fool the expert, it is safe to say that the learned features are effective.
However, the original GANs cannot directly handle the energy disaggregation problem. At first, energy disaggregation is not a generative problem. It is a BSS problem. Secondly, the discriminator for energy disaggregation should tell the signal sources, i.e., corresponding appliances, instead of fake and real. 
To address the above issues, we formulated AED from the following two aspects: 
1) The generator is formulated as a feature representation network. Since different appliances share the same observation from the mains readings, we deploy a shared convolutional neural network to learn the temporal features of appliances.
2) Considering the complex multimode signal structures of different appliances, we train multiple discriminators to enable precise and fine-grained source separation. As a result, we formulate our AED as a multi-adversarial learning model which addresses BSS problem. 

{Based on this design, we now sketch how our method can be applied in real-world scenarios, where variant appliances and houses exist. First, our model is trained on the large-scale supervised datasets, which inherently reduces the over-fitting risk and better generalizes to unseen houses in the wild to a certain extent. However, like other data-driven algorithms, training model on a specific target appliance inevitably results in a generator that is appliance-biased, i.e., the model learns features that benefit the current appliance while ineffectual for other appliances. Our method alleviates this issue by a multi-adversarial learning paradigm, which aims to exploit the common information across multiple appliance-specific features encoded by multiple feature extractors of appliances available at hand, hoping the learnt features are effective for other unseen appliances in real-world without further fine-tuning. It is worth noting that although training a deep model is computationally expensive, it does not need to be performed very often, i.e., once the model is trained, we only need the aggregated data and a lightweight forwarding inference to obtain the disaggregated energy consumption for each appliance, which is much fast and more economic than intrusive load monitoring approaches such as deploying smart sensor devices (e.g., smart plugs) for each appliance. As pointed out by Kelly \textit{et al.} \cite{kelly2015neural} who first suggest applying deep neural networks to energy disaggregation, the inference could be performed on a compact compute device within each house without GPUs, and the energy cost of DNN inference is also marginal in comparison to the saved energy. On the contrary, smart plugs are expensive to install and hard to maintain simultaneously. In a nutshell, the main contributions of this work can be summarized as follows:}
\begin{enumerate}[1)]
\item We propose a novel NILM method named adversarial energy disaggregation (AED). Different from previous DNN methods which use conventional CNNs and RNNs, AED introduces adversarial learning into the energy disaggregation problems. Experiments verify that AED can significantly outperform previous state-of-the-art algorithms. 

\item We report a new CNN structure (reported in Fig.~\ref{fig:generator}) which is able to learn discriminative feature representations for different appliances. At the same time, we present the multi-adversarial BSS framework (reported in Fig.~\ref{fig:idea}) which enables fine-grained single-channel blind source separation.

\item Energy efficiency is a great challenge facing humanity in the~21st century. Our method develops a practical solution for NILM which takes steps towards energy efficiency.
\end{enumerate}

\textcolor{black}{The rest of this paper is organized as follows. Section 2 makes a review of previous work related to this paper. Section 3 presents the notations and elaborates on our proposed framework. In Section 4, we conduct extensive experiments on two widely used energy disaggregation datasets and show the superior of our method compared to existing methods. Finally, we make a conclusion and discussion in Section 5.}

\section{Related Work}
\subsection{Energy Disaggregation}
Energy Disaggregation, which is also known as non-intrusive load monitoring (NILM), was first introduced by Hart et al.~\cite{hart1992nonintrusive}. The task of energy disaggregation is to separate individual signals of different appliances from an overall observation. In previous work, a battery of methods have been explored to address the energy disaggregation problem. According to~\cite{shin2019subtask}, the most popular model for NILM is the Factorial Hidden Markov Model (FHMM)~\cite{ghahramani1996factorial}. For instance, Zhong et al.~\cite{zhong2014signal} proposed an additive FHMM with signal aggregate constraints. Shaloudegi et al.~\cite{shaloudegi2016sdp} developed a scalable approximate inference FHMM algorithm based on a semi-definite relaxation combined with randomized rounding. Apart from FHMM, other techniques such as sparse coding~\cite{kolter2010energy} and support vector machines
(SVMs) were also leveraged in the community~\cite{faustine2017survey}. 
Recently, deep neural networks have been introduced into energy disaggregation~\cite{kelly2015neural,zhang2018sequence,shin2019subtask,bejarano2019deep}. 
\textcolor{black}{Due to the superior performance in time series processing, Recurrent Neural Networks (RNN) and its variants such as Long Short-Term Memory Networks (LSTM) and Gated Recurrent Units (GRU) have been employed in NILM. For instance, Mauch et al.~\cite{mauch2015new} advocate to use multiple bidirectional LSTM layers to tackle the problem of energy disaggregation. Kim et al. \cite{kim2017nonintrusive} address the energy disaggregation by an LSTM model as well as a novel signature to boost the performance. Kaselimi et al. \cite{kaselimi2020context} propose the CoBiLSTM, which employs the representational power of LSTM networks and realizes the adaptation to the external environment.}

\textcolor{black}{While the majority of studies that model time series data is to leverage RNNs, convolutional neural networks (CNNs) has also been adapted to the context of energy disaggregation due to its powerful ability in extracting local features of patterns. To alleviate the computational complexity issue in long input sequences, a common trick is to use a sliding window instead of the whole sequence. For instance,} Kelly et al.~\cite{kelly2015neural} investigated several DNNs, such as convolutional neural networks, recurrent neural networks and denoising autoencoders, to handle the energy disaggregation problem, \textcolor{black}{they propose the sequence-to-sequence learning with slide windows and show the superior of deep learning methods to traditional methods. Chen et al. \cite{chen2018convolutional} propose a convolutional sequence-to-sequence model and introduce the gated linear unit (GLU) convolutional blocks into energy disaggregation, which are used to extract information from the main readings and control the features output by conventional CNN layers.} Zhang et al.~\cite{zhang2018sequence} further proposed a sequence-to-point (seq2point) network that only predicts~the~middle~point of the window, which significantly improves the performance of previous SOTA methods. 

\subsection{Deep Adversarial Networks}
The most popular deep adversarial learning paradigm is generative adversarial networks (GANs)~\cite{goodfellow2014generative}, which consist of two neural networks known as generator and discriminator respectively. In the training process, the generator synthesizes fake samples to fool the discriminator. At the same time, the discriminator tries its best to distinguish the fake from the real. 

\textcolor{black}{It is worth noting that some studies have been carried out with the GANs in the literature. As an early attempt to leverage GANs in energy disaggregation, Bao et al. \cite{bao2018enhancing} propose to integrate the generator of a pre-trained GAN into the conventional Neural NILM process to generate the appliance load sequences more accurately. Later, Kaselimi et al. \cite{kaselimi2020energan} use a CNN based seeder and generator to encode the main reading signals and produce the appliance consumption respectively, with adversarial learning, the produced load sequences would match the ground truth as close as possible. Recently, Pan et al. \cite{pan2020sequence} propose a sequence-to-subsequence learning in NILM, which makes a trade-off between sequence-to-sequence and sequence-to-point learning and predicts a subsequence of the main window. To this end, conditional GAN \cite{mirza2014conditional} is used to encourage the generator to produce the appliance load sequences conditioned on the input main reading windows.}

In our model, although we share the similar spirit of GANs and tailor a model for energy disaggregation, \textcolor{black}{our proposed method is significantly different from previous work in both motivation and formulation.} Specifically, we leverage the idea of adversarial learning and report new contributions to both generator and discriminator. On one hand, we train multiple discriminators, one for each appliance, to capture the complex multimode structures. On the other hand, we propose a new deep network structure to learn the feature representations of different appliances. {Furthermore, previous studies all use GAN for generating the whole appliance load sequences, i.e., they employ an end-to-end learning model and generate the final predictions directly, the adversarial process is carried out on the final output space. However, our model can be decomposed into a feature extractor (generator) which learns a latent feature space that effective for energy disaggregation, and a predictor (linear layer) that predicts the energy consumption based on the learnt latent features, the adversarial learning is carried out on the latent features output by the feature extractor.} Thus, the reported work is significantly different from our AED.

\section{Adversarial Energy Disaggregation}
\textcolor{black}{In this section, we present our proposed method in detail. For a better understanding, we first review and formulate the problem of energy disaggregation. We present some notations and definitions that used in our method. After that, we introduce the detail framework of our proposed adversarial energy disaggregation.}

\subsection{Problem and Formulation Overview}
\begin{figure*}[t]
  \begin{center}
  \includegraphics[width=1\linewidth]{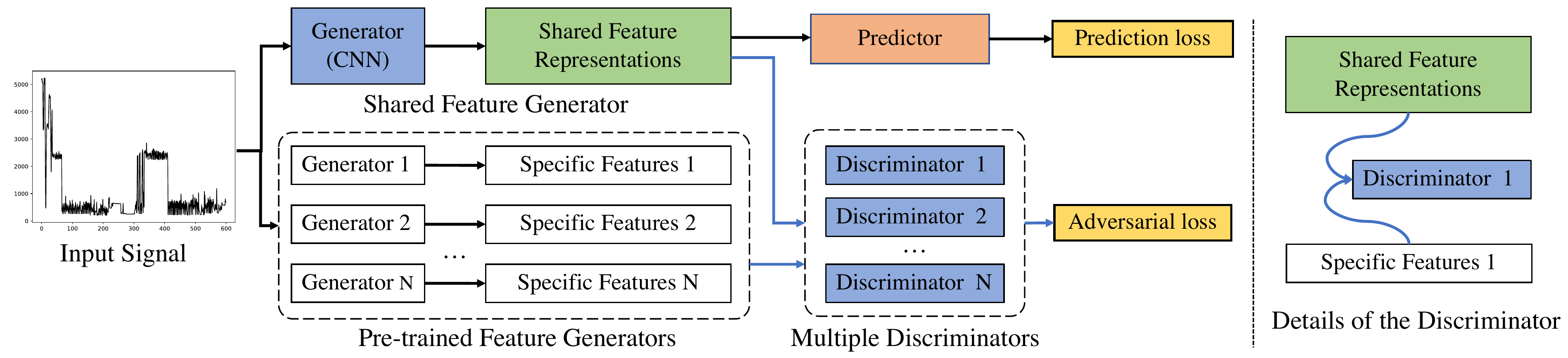}
  \end{center}
  \caption{Idea illustration. The left sub-figure shows the main idea of our method. The main architecture consists of a feature generator, a predictor and $N$ appliance discriminators, where $N$ is the total number of appliances. During training, the appliance discriminator distinguishes the shared features from the appliance-specific features. Once a discriminator is confused, we assume that the shared features capture the latent structures of the corresponding specific features. For conciseness, the left figure is simplified. The right sub-figure shows the inputs of a discriminator. Here we take the discriminator~$1$ as an example. The rest can be done in the same manner. The appliance-specific feature generators, which have the same network structure with the shared generator, are pre-trained. More details about the networks are reported in the context.}
  \label{fig:idea}
  \end{figure*}

The goal of energy disaggregation is to recover the energy consumption of individual appliances from the mains power readings, which measure the whole-home energy consumption. Recent advances in energy disaggregation~\cite{zhang2018sequence,kelly2015neural} reveal that different appliances can be distinguished from the mains readings by learning deep features. Since energy disaggregation is an unidentifiable BSS problem, the signals of different appliances are mixed together in the mains readings. In the test process, we only have the aggregated signals for feature learning. Thus, the main challenge of feature-learning-based energy disaggregation methods is whether the learned features can capture the multimode structures of different appliances. For convenience, we refer the features learned from the aggregated signals and appliance-specific signals as shared features and specific features, respectively. The specific features can identify the corresponding appliances. In our model, we encourage the shared features to capture the characteristics of each appliances. To this end, we propose the multi-adversarial learning approach as illustrated in Fig.~\ref{fig:idea}. Specifically, we first train multiple appliance-specific feature generators, one for each appliance, to learn the appliance-specific features. During the adversarial learning, the shared generator learn features to confuse the discriminators. Once the discriminators are confused, it is assumed that the shared feature representations have captured all the multimode structures of different appliances. In addition, a predictor (classifier) is also trained on the shared features to leverage the supervised information. In the following subsections, we report the details of our proposed method.

\subsection{Notations and Definitions}
Suppose there are $N$ appliances in a household, and we observed the mains readings $Y$ that represents the aggregate power of all appliances in Watts, where $Y=\left(y_{1}, y_{2}, \dots, y_{T}\right)$ and $Y(t)$ denotes the main readings at time $t$. For the $i$-th appliance, its power consumption is represented by a sequence $X_{i}=\left(x_{i 1}, x_{i 2}, \ldots, x_{i_T}\right)$ and $X_{i}(t)$ denotes the assumption of $i$-th appliance at time $t$. The relationship between $X$ and $Y$ can be represented by $Y(t)=\sum_{i=1}^{N} X_{i}(t)+ \epsilon(t)$, where $\epsilon(t)$ is the random noise which follows a Gaussian with mean 0 and variance $\sigma_{t}^{2}$, i.e., $\epsilon(t) \sim \mathcal{N}\left(0, \sigma_{t}^{2}\right)$. The  task of energy disaggregation is to infer the individual consumptions of each appliance, i.e., $X_{i}=\left\{X_{i}(t)\right\}_{t=1}^{T}$, according to the mains readings $Y$. In our proposed deep neural networks, we use letters $G$, $D$ and $C$ to denote the generator, the discriminator and the predictor, respectively. 

\subsection{The Generators for Feature Representation}
\begin{figure*}[t]
  \begin{center}
  \includegraphics[width=1\linewidth]{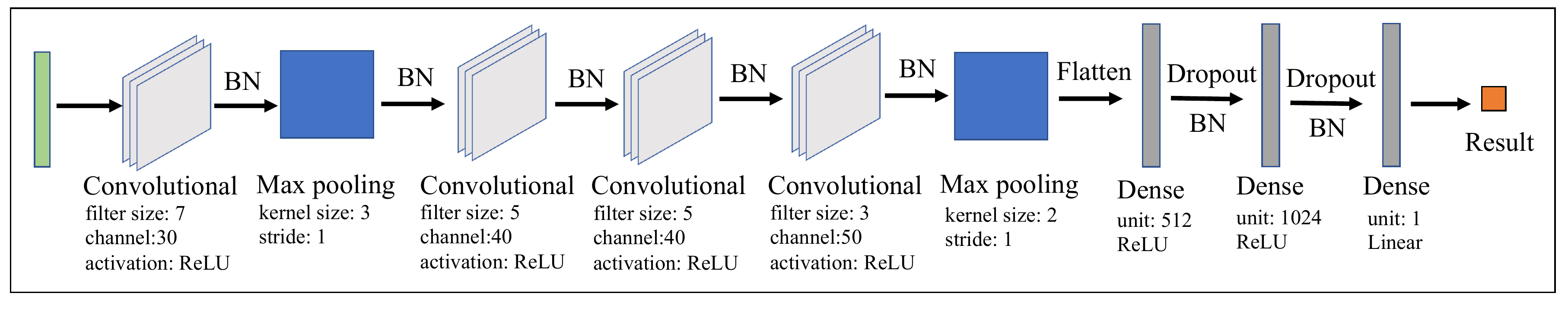}
  \end{center}
  \caption{An illustration of the structure and implementation details of the feature generator (the four convolutional layers and two max pooling layers) and the predictor (the three dense layers).}
  \label{fig:generator}
  \end{figure*}

The feature learning process in deep energy disaggregation can be formulated as a sequence-to-sequence (seq2seq)~\cite{kelly2015neural} or sequence-to-point (seq2point)~\cite{zhang2018sequence} learning problem. Theoretical analysis and experimental evaluation show that seq2point has better performance~\cite{zhang2018sequence}. Specifically, seq2point trains a neural network to predict the midpoint power of an appliance when giving a window of the mains readings as the input. In this paper, we follow the paradigm of seq2point learning. The feature learning network $G$ (the generator of our adversarial model, parameterized by $\theta_G$) takes a mains window $Y_{t:t+W-1}$ as input, and outputs the midpoint power $x_{\tau}$ of the corresponding window $X_{t: t+W-1}$ of the target appliance, where $W$ is the window size and $\tau = t + \lfloor W / 2\rfloor$. The mapping can be described as $x_{\tau}=G \left(Y_{t: t+W-1}\right)+\epsilon$, and the loss function of this problem can be formulated as follows:
\begin{equation}
\begin{array}{c}
\mathcal{L}_G=\sum_{t=1}^{T-W+1} \log p\left(x_{\tau} | Y_{t: t+W-1}, \theta_{G}\right),
\end{array}
\label{eq:s2p}
\end{equation}
where $\theta_{G}$ denotes the parameters of network $G$. This formulation is effective because it can make full use of the state information of the main readings before and after the midpoint time to predict the power of the target appliance at the specific time. 
Theoretically, the problem in~\eqref{eq:s2p} can be maximized by many architectures, such as denoising autoencoders, convolutional neural networks (CNN) and recurrent neural networks (RNN). In this paper, we propose a new CNN structure for energy disaggregation as shown in Fig.~\ref{fig:generator}. The proposed feature learning network consists of four convolutional layers and two max pooling layers. It is worth noting that the signal at time $t$ is generally similar to the signal at $t-\varepsilon$ and $t+\varepsilon$, where $\varepsilon$ is a small number. Therefore, the max pooling layers in our CNN can be used to avoid over-fitting and improve the generalization ability.

As illustrated in Fig.~\ref{fig:idea}, our model pre-trains multiple appliance-specific generators, one for each appliance, to learn the appliance-specific features. To this end, we also train a predictor $C$ (parameterized by $\theta_C$) which consists of three fully connected layers to leverage the supervised ground truth information. For a given appliance $X_i$ and its ground truth information at time $t$, i.e, $x_{it}$, we deploy the following loss to train the appliance-specific generator $G_i$ and the predictor $C_i$ via an end-to-end manner:
\begin{equation}
\begin{array}{c}
  \mathcal{L}_\mathrm{pred}\left({X}_{i}, {Y}; G_{i}, C_{i}\right)= \\
  \mathbb{E}_{\left(x_{i t}, y_{t\!-\!W/2:t+W/2}\right) \sim\left({X}_{i}, {Y}\right)} || x_{i t} \!-\! C_{i}\left(G_{i}\left(y_{t\!-\!W/2:t+W/2}\right)\right)||_2^{2}.
  \end{array}
\end{equation}
It is worth noting that the appliance-specific generators $G_i$ ($i \in [1,N]$) share the same structure, as shown in Fig.~\ref{fig:generator}, with the shared generator $G$. During the adversarial learning process, the pre-trained generators will be fixed to extract appliance-specific features.

\subsection{Adversarial Energy Disaggregation}



In the test stage, we only have the aggregated signal which consists of unidentified appliances, e.g., we didn't know which appliance contributes to the overall mains reading. Thus, it is necessary to learn a shared feature generator which is able to capture the multimode structure of individual signals.

In order to obtain a more generalized feature space, we try to find some latent common information from the load features encoded by multiple extractors of each appliances. Technically, we propose to use multiple-adversarial learning as illustrated in Fig.~\ref{fig:idea}. During the training process, we firstly use the $N$ pre-trained extractors to extract $N$ features for each mains readings window, then we train a generator $G$ to compete against all the discriminators simultaneously. As the adversarial process continuing, the generator will gradually learn to extract shared features and finally fool all the $N$ discriminators~\cite{goodfellow2014generative}. Consequently, the feature representations learned in this way will be able to capture the multimode structure which embedded in the multiple appliance-specific feature spaces extracted by $N$ pre-trained generators. The multi-adversarial learning process can be written as:
\begin{equation}\begin{array}{c}
  \mathcal{L}_\mathrm{adv}({Y}  ; G, D_{1}, D_{2}, \ldots, D_{N})= \\
  \sum_{i=1}^{N}(\mathbb{E}_{y_{t-W/2:t+W/2} \sim {Y}}[\log (D_{i}(G(y_{t-W/2:t+W/2}))) + \\  \log\! (1-D_{i} (G_{i} (y_{t-W/2:t+W/2}) ))]),
  \end{array}\end{equation}    
where $D_{i}$ represents the $i$-th discriminator that aims to distinguish whether the features come from shared generator $G$ or the $i$-th pre-trained extractor $G_i$. After several rounds of training, the generator $G$ would be able to extract shared feature representations for all the appliances. Figuratively speaking, features learned by the shared generator can be seen as ``fake'' to fool the discriminator, while features learned by the appliance-specific generators can be regarded as ``real''. Once the discriminators are confused, the shared generator captures the complex multimode structures of each appliance.




At last, since the adversarial model is required to decompose the mains rather than only learning feature representations, we further train a predictor (classifier) in the adversarial learning framework. As a result, we have the overall formulation:
\begin{equation}
  \min_{G,C } ~~\max _{D_{1}, D_{2}, \ldots, D_{N}} \mathcal{L}_\mathrm{AED}= \\
\mathcal{L}_\mathrm{adv} + \lambda \mathcal{L}_\mathrm{pred},
\end{equation}
where $C$ denotes the classifier and $\lambda >0$ denotes the trade off parameter. It is easy to see that the whole training process of our proposed AED is via an adversarial fashion. On one hand, the generator $G$ and classifier $C$ are trained together to minimize prediction loss and multi-adversarial domain generalization loss. On the other hand, the domain discriminator $D_{1}, D_{2}, \ldots, D_{N}$ are trained together to maximize the multi-adversarial domain generalization loss so as to compete~with~the~generator. 

\begin{algorithm}[t]
  \caption{Adversarial Energy Disaggregation}\label{algorithm}
  \begin{algorithmic}[1]
  \REQUIRE 
   Sequence $X$ of the target appliance, main reading sequence $Y$, $N$ pre-trained appliance-specific generators $G_1$, $G_2$, ..., $G_N$, $N$ discriminators $D_1$, $D_2$, ..., $D_N$, trainable target generator $G$ and predictor $C$, hyperparameter $\lambda_{cls}$, batch size $B$, maximum epoch $M$.
  \ENSURE
  Optimal $\theta_{G}$, $\theta_{C}$ for the target model $G$ and $C$.
  \FOR{epoch $\gets$ 1 to $M$}
  \FOR{batch $\gets$ 1 to $K$}
  \STATE $g_G \leftarrow 0$
  \STATE Sample $B$ main windows $\{y_{i}\}_{i=1}^{B} \sim Y$ and obtain the corresponding $x_i$ of each $y_{i}$; 
  \STATE Extract the features for $\{y_{i}\}_{i=1}^{B}$ via $\{G(y_i)\}_{i=1}^{B}$;
  \FOR{ $j$ $\gets$ 1 to $N$}
  \STATE Extract the appliance-specific features for $\{y_{i}\}_{i=1}^{B}$ via $ \{G_j(y_i)\}_{i=1}^{B}$
  \STATE Update $D_j$ via $\theta_{D_j} \leftarrow \operatorname{Adam} \left(\nabla_{\theta_{D_j}} \left( \mathcal{L}_{adv} \right)\right)$ according to Eq. (3).
  \STATE $g_G += \nabla_{\theta_{G}} \left( -\mathcal{L}_{adv} \right)$
  \ENDFOR
  \STATE Calculate $\mathcal{L}_{pred}$ according to Eq. (2). 
  \STATE Update $G$ via $\theta_{G} \leftarrow \operatorname{Adam} \left(g_G + \nabla_{\theta_{G}} \left(\lambda \mathcal{L}_\mathrm{pred} \right)\right)$
  \STATE Update $C$ via $\theta_{C} \leftarrow \operatorname{Adam} \left(\nabla_{\theta_{C}} \left( \lambda \mathcal{L}_\mathrm{pred} \right)\right)$
  \ENDFOR 
  \ENDFOR
  \end{algorithmic}
\end{algorithm}

\section{Experiments}
In this section, we verify the proposed method on two real-world datasets collected from US and UK families. We compare AED with several previous state-of-the-art approaches~which~deploy different techniques. Our model is implemented by PyTorch and trained on NVIDIA GTX 2080Ti GPUs. The datasets used in this paper can be downloaded via the provided links reported in the~dataset description section. Our code and data are released at {\it https://github.com/lijin118/}. 

\subsection{Datasets}
We testify our AED on two popular datasets for energy disaggregation. The description is listed as follows.

{\bf REDD}~\cite{kolter2011redd} dataset~\footnote{http://redd.csail.mit.edu/} is a widely used benchmark for NILM tasks. The dataset records the domestic energy consumption, at both appliance-level and whole-house level, of six US houses from November 2012 to January 2015. The recording intervals of the appliance and mains readings are 3 seconds and 1 second, respectively. Following previous work~\cite{kelly2015neural,zhang2018sequence}, we use houses 2 to 6 for training and house 1 for test. \textcolor{black}{For the same reasons to those in \cite{zhang2018sequence}, we only consider microwave, fridge, dish washer and washing machine in this paper.}

{\bf UK-DALE}~\cite{kelly2015uk} dataset \footnote{https://jack-kelly.com/data/} records both appliance-level and whole-house level energy consumption of five UK houses from November 2012 to January 2015. The records are read in every 6 seconds. In this paper, we follow the previous work~\cite{zhang2018sequence} and choose washing machine, kettle, microwave, dish washer and fridge for evaluations. Houses 1, 3, 4 and 5 are used for training and house 2 used for test.

\begin{table}[]
  \caption{The parameters used in this paper for processing the data. The length unit is point and the power unit is Watt.}
  \label{tab:normalization}
  \begin{tabular}{@{}lcccc@{}}
  \cmidrule(r){1-4}
  Appliance       & Window length & Mean & Standard deviation &  \\ \cmidrule(r){1-4}
  Aggregate       & 599           & 522  & 814                &  \\
  Kettle          & 599           & 700  & 1000               &  \\
  Microwave       & 599           & 500  & 800                &  \\
  Fridge          & 599           & 200  & 400                &  \\
  Dishwasher      & 599           & 700  & 1000               &  \\
  Washing machine & 599           & 400  & 700                &  \\ \cmidrule(r){1-4}
  \end{tabular}
  \end{table}

\subsection{Implementation Details}
\textcolor{black}{\textbf{Network architecture.}} Our model consists of three main components: the generator $G$, the discriminator $D$ and the predictor $C$. The implementation details of $G$ and $C$ are reported in Fig.~\ref{fig:generator}. \textcolor{black}{The generator $G$ consists of four convolutional layers and two max pooling layers. Specifically, these two max pooling layers are set after the first and the last convolutional layers and have pool sizes of 3 and 2, respectively. The filter size and channel of the convolutional layers are set to \{7 $\times$ 1, 5 $\times$ 1, 5 $\times$ 1, 3 $\times$ 1\} and \{30, 40, 40, 50\}, respectively. Replication pad is used to make the sequence length invariant after convolutional operation.} The discriminator $D$ is implemented by three fully connected (FC) layers, i.e., FC-ReLU-FC-ReLU-FC-Sigmoid. \textcolor{black}{We adopt batch normalization and dropout in this paper.} The pre-trained $G_i$ share the same structure with $G$. Different discriminators also have the same network architecture. 

\textcolor{black}{\textbf{Hyper-parameter setting.}} We optimize our networks by Adam~\cite{kingma2014adam} optimizer and the parameters are $\beta_1$= 0.9, $\beta_2=0.999$ and $\epsilon=10^{-8}$. The learning rate is $0.001$. The window size is $599$ (a sample window contains $599$ recording points) and the batch size is $1000$. We set the number of maximum epoch to $50$. The hyper-parameter $\lambda=0.05$. 

\textcolor{black}{\textbf{Data Pre-processing.}} For fair comparisons, we follow previous work~\cite{zhang2018sequence} to pre-process data. \textcolor{black}{Specifically, we first align the main readings with the appliance readings by timestamps in each house then concatenate them together. After that, we perform a normalization on the raw data by
\begin{equation}
  \frac{x^{i}_{t}-\bar{x}^{i}}{\sigma^{i}}
  \end{equation}
where $x^{i}_{t}$ denotes the power reading at time $t$ of the $i$-th appliance. $\bar{x}^{i}$ and $\sigma^{i}$ stand for the mean value and the standard deviation of the $i$-th appliance, respectively. The specific parameters for normalization can be found in Table~\ref{tab:normalization}. The normalized data is then fed into the model for training.}  


\renewcommand\arraystretch{1.2} 
\begin{table*}
  \caption{Quantitative results on REDD dataset. Best results are highlighted in bold. For a better layout, we use {\it Microwv, Dishwsh, Washmch} to denote {\it Microwave oven, Dishwasher, Washingmachine}.}
  \label{tab:redd}
  \centering
  \begin{tabular}{llccccc}
    \toprule
    Metric & Methods & Microwv & Fridge & Dishwsh & Washmch  & Average  \\
    \midrule
    \multirow{4}*{MAE} & AFHMM~\cite{kolter2012approximate} & {\bf 11.85} & 69.80 & 155.25  & 14.25 & $62.78 \pm 58.20$  \\
    \cmidrule(r){2-7}
                     & CO~\cite{batra2014nilmtk} & 62.85 & 78.50 & 108.24 & 90.63 & $85.06 \pm 37.25$\\
    \cmidrule(r){2-7}
                     & FHMM~\cite{batra2014nilmtk} & 71.12 & 89.67 & 99.79 & 65.77 & $81.59 \pm 42.14$\\
    \cmidrule(r){2-7}
                     & SGN~\cite{shin2019subtask} & 17.52 & \bf{23.89} & 14.97 & 20.07 & $19.09$\\
    \cmidrule(r){2-7}
                     & Seq2seq~\cite{zhang2018sequence} & 33.27 & 30.63 & 19.45 & 22.86 & $26.55 \pm 5.61$\\
    \cmidrule(r){2-7}
                      & Seq2point~\cite{zhang2018sequence} & 29.602 & 34.118 & 22.476  &16.130 & $24.24 \pm 7.45$ \\
    \cmidrule(r){2-7}
                     & AED [Ours] & 17.914 &  29.770 & {\bf 11.527} & {\bf 11.533} & ${\bf 17.68 \pm 7.44}$\\
     \midrule
     \multirow{4}*{SAE} & AFHMM~\cite{kolter2012approximate} & 0.84 & 0.99 & 7.19  &0.07 & $2.28 \pm 2.86$ \\
     \cmidrule(r){2-7}
                      & CO~\cite{batra2014nilmtk} & 3.48 & 0.11 & 4.56 & 2.77 & $ 2.73 \pm 1.53 $ \\
     \cmidrule(r){2-7}
                      & FHMM~\cite{batra2014nilmtk} & 3.07 & 0.37 & 3.53 & 1.11 & $ 1.77 \pm 0.87 $ \\
    \cmidrule(r){2-7}
                      & SGN~\cite{shin2019subtask} & - & - & - & - & -\\
    \cmidrule(r){2-7}
                     & Seq2seq~\cite{zhang2018sequence} & 0.24 & 0.11 & 0.56 & 0.51 & $ 0.36 \pm 0.183 $ \\
    \cmidrule(r){2-7}
                      & Seq2point~\cite{zhang2018sequence} & {\bf 0.037} & 0.100 & 0.701 & 0.240 & $0.27 \pm 0.26$\\
    \cmidrule(r){2-7}
                     & AED [Ours] & 0.171 & {\bf 0.079} & {\bf 0.339} & {\bf 0.029} & ${\bf 0.15 \pm 0.11}$\\
     \bottomrule
  \end{tabular}
\end{table*}

\begin{table*}
  \caption{Quantitative results on UK-DALE dataset. Best results are highlighted in bold.}
  \label{tab:dale}
  \centering
  \resizebox{\textwidth}{!}{
  \begin{tabular}{llcccccc}
    \toprule
    Metric & Methods & Kettle & Microwv & Fridge & Dishwsh & Washmch  & Average  \\
    \midrule
    \multirow{5}*{MAE} & AFHMM~\cite{kolter2012approximate} & 47.38 & 21.28 & 42.35  &199.84 & 103.24 & $89.79 \pm 64.50$  \\
    \cmidrule(r){2-8}
                       & CO~\cite{batra2014nilmtk}  & 47.65 & 100.26 & 78.76 & 109.96 & 91.52 & $85.63 \pm 60.20$\\ 
    \cmidrule(r){2-8}
                       & FHMM~\cite{batra2014nilmtk} &  45.23 & 46.08 & 57.81 & 50.18 & 70.13 & $54.09 \pm 36.33$\\
    \cmidrule(r){2-8}
                       & SGN~\cite{shin2019subtask} &  7.08 & 6.26 & 15.79 & 15.50 &  12.31 & $11.39$\\
    \cmidrule(r){2-8}
                       & Seq2subseq (N-I+U)~\cite{pan2020sequence} &  10.34 & {\bf 5.77} & 35.44 & 28.68 & 20.65 & $20.18$\\
    \cmidrule(r){2-8}
                      & Seq2seq~\cite{kelly2015neural} & 13.000 & 14.559 & 38.451  & 237.96 & 163.468 & $93.49 \pm 91.11$\\
    \cmidrule(r){2-8}
                      & Seq2point~\cite{zhang2018sequence} & 8.656 & 8.700 &20.894 & 29.724 & 12.724 & $16.14 \pm 8.12$\\
    \cmidrule(r){2-8}
                      & AED [Ours] &{\bf 5.645} &  5.948 & {\bf 13.324} & {\bf 15.062} &  {\bf 5.764} & ${\bf 9.15 \pm 5.72}$\\
      \midrule
      \multirow{5}*{SAE} & AFHMM~\cite{kolter2012approximate} & 1.06 & 1.04 & 0.98  &4.50 & 8.28 & $3.17 \pm 2.88$  \\
      \cmidrule(r){2-8}
                         & CO~\cite{batra2014nilmtk}  & 1.60 & 11.81 & 1.22  & 1.83 & 2.04 & $3.70 \pm 2.24$  \\
      \cmidrule(r){2-8}
                         & FHMM~\cite{batra2014nilmtk} & 8.77 & 4.60 & 0.34  & 0.39 & 1.74 & $3.17 \pm 2.02$  \\
      \cmidrule(r){2-8}
                         & SGN~\cite{shin2019subtask} & - & - & -  & - & - & -  \\
      \cmidrule(r){2-8}
                         & Seq2subseq (N-I+U)~\cite{pan2020sequence} & 0.081 & 0.365 & 0.477  & 0.709 & 0.317 & $0.390$  \\
      \cmidrule(r){2-8}
                         & Seq2seq~\cite{kelly2015neural} & 0.085 & 1.348 & 0.502  &4.237 & 13.831 & $4.00 \pm 5.12$\\
      \cmidrule(r){2-8}
                         & Seq2point~\cite{zhang2018sequence} & 0.072 & 0.430 & 0.121 & 0.376 & 0.245 & $0.25 \pm 0.14$\\
      \cmidrule(r){2-8}
                         & AED [Ours] & {\bf 0.046}  & {\bf 0.146} & {\bf 0.038} & {\bf 0.163} & {\bf 0.053} & ${\bf 0.09 \pm 0.11}$  \\
      \bottomrule
  \end{tabular}}
\end{table*}

\subsection{Compared Methods and Evaluation Metrics}
We compare our method with several previous representative approaches which deploy different techniques. Specifically, AFHMM~\cite{kolter2012approximate} is a traditional method which is built on additive factorial hidden Markov models. Seq2seq~\cite{kelly2015neural, zhang2018sequence} is a deep method which leverages sequence to sequence learning. Seq2point~\cite{zhang2018sequence} is a recently reported deep methods which deploys sequence to point learning. {Seq2subseq~\cite{pan2020sequence} is a trade-off method between Seq2seq and seq2point, which employs conditional GAN to learn the output distribution for each appliance. We cite the results of the batch normalization version (i.e., N-I-U) from \cite{pan2020sequence} for a fair comparison. Subtask Gated Networks (SGN) \cite{shin2019subtask} explicitly considers the on/off states of appliances and combines two subtask networks (i.e., a regression network and a classification network) to make final predictions. In addition, we also compare our method with two benchmark algorithms implemented in the publicly available NILM toolkit (NILMTK) \cite{batra2014nilmtk} platform, namely combinatorial optimization (CO) and factorial hidden Markov model (FHMM). The experiments of CO and FHMM are conducted based on the rewrote experiment API of NILMTK provided by Batra \textit{et al.} \cite{batra2019towards} under the same setting as other algorithms.} It is worth noting that the reported results of seq2point are the best results we can achieve by running the authors' code. For fair comparisons, we follow~\cite{zhang2018sequence} and use the following two metrics: 
 \renewcommand\arraystretch{1.5} 
 \begin{equation}
 \begin{array}{l}
  \mathrm{MAE}=\frac{1}{T}\sum_{t=1}^T |\hat{x_t}-x_t|, \\ \mathrm{SAE}=\frac{|\hat{r}-t|}{r},
  \end{array}
\end{equation}
\renewcommand\arraystretch{1} 
where $\hat{x}$, $x$, $\hat{r}$, and $r$ denote the predicted consumption of an appliance at time $t$, the ground truth consumption of an appliance at time $t$, predicted total energy consumption of an appliance, and the ground truth total consumption of an appliance, respectively. It is easy to observe that $r=\sum_t x_t$ and $\hat{r}=\sum_t \hat{x}_t$. It is worth noting that the used MAE and SAE have no linear relationship. The mean absolute error (MAE) reflects the fine-grained performance of prediction at every recording point. The normalized signal aggregate error (SAE) reports a more global prediction accuracy. 

\subsection{Quantitative Results}
The quantitative results on REDD and UK-DALE are reported in Table~\ref{tab:redd} and Table~\ref{tab:dale}, respectively. From the results in Table~\ref{tab:redd}, we can see that our method achieves the best results in two out of four appliances. In terms of average performance, our AED has both the smallest MAE and SAE. The average MAE of our AED is $17.68$, while the previous baseline method seq2point achieves only $24.24$. The improvement is $27\%$. We can also notice that our method significantly reduces the MAE of dishwasher by $48.7\%$ compared to seq2point which also uses the sequence-to-point architecture. For other appliances like microwave and washing machine, our method can reduce the MAE value by $39.5\%$ and $28.5\%$, respectively. Furthermore, our method also decreases the average SAE by $44.4\%$. The results verify that our method has a much better NILM performance with respect to both timely prediction (MAE) and overall prediction (SAE). 

\begin{figure*}[t]
  \begin{center}    
  \includegraphics[width=0.99\linewidth]{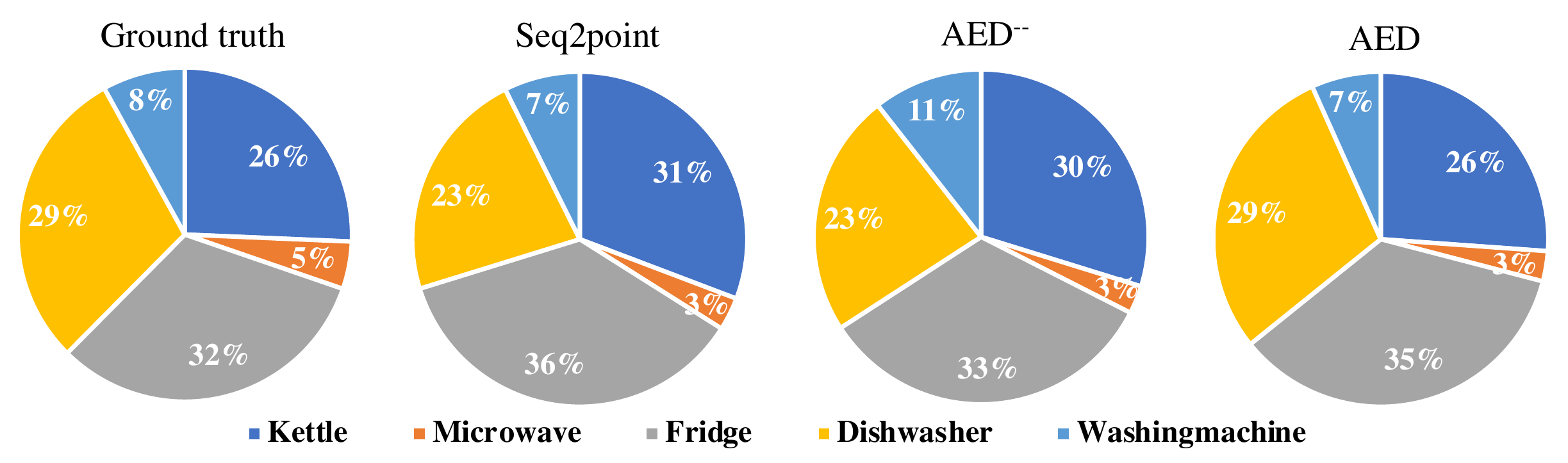}    
  \end{center}
  \caption{Percentages of total energy usage in the test set of UK-DALE. From left to right, the four figures show the results of ground truth, seq2point, AED without multi-adversarial learning and AED.}
  \label{fig:pie}
  \end{figure*}

The results in Table~\ref{tab:dale} draw the similar conclusions. {Our AED outperforms the previous state-of-the-art methods in four out of five appliances.} Compared with seq2point, our AED reduces the average MAE and SAE by $43.3\%$ and $64.0\%$, respectively. Specifically, the MAE values of kettle, microwave, fridge, dishwasher and washing machine are reduced by percents of $34.8\%$, $31.6\%$, $36.2\%$, $49.3\%$ and $54.7\%$, respectively. {Comparing with Seq2subseq, our method shows significant improvements on kettle, fridge, dishwasher and washing machine and achieves a comparable result on microwave with respect to MAE.} It is also worth noting that our method has the smallest standard deviations, which indicates that our method is robust for different appliances. Our method trains the feature representation network by multi-adversarial learning. Thus, the multimode structures embedded in different appliance-specific signals can~be~preserved, which explains the small standard deviation of our results. At last, we report the percentage of total energy consumption of different methods in Fig.~\ref{fig:pie}. The results give a straightforward report of the family energy usage in terms of different device types. We can see that our AED is able to generate a report which is very close to the ground truth.

\begin{figure*}[t]
  \begin{center}
  \includegraphics[width=1\linewidth]{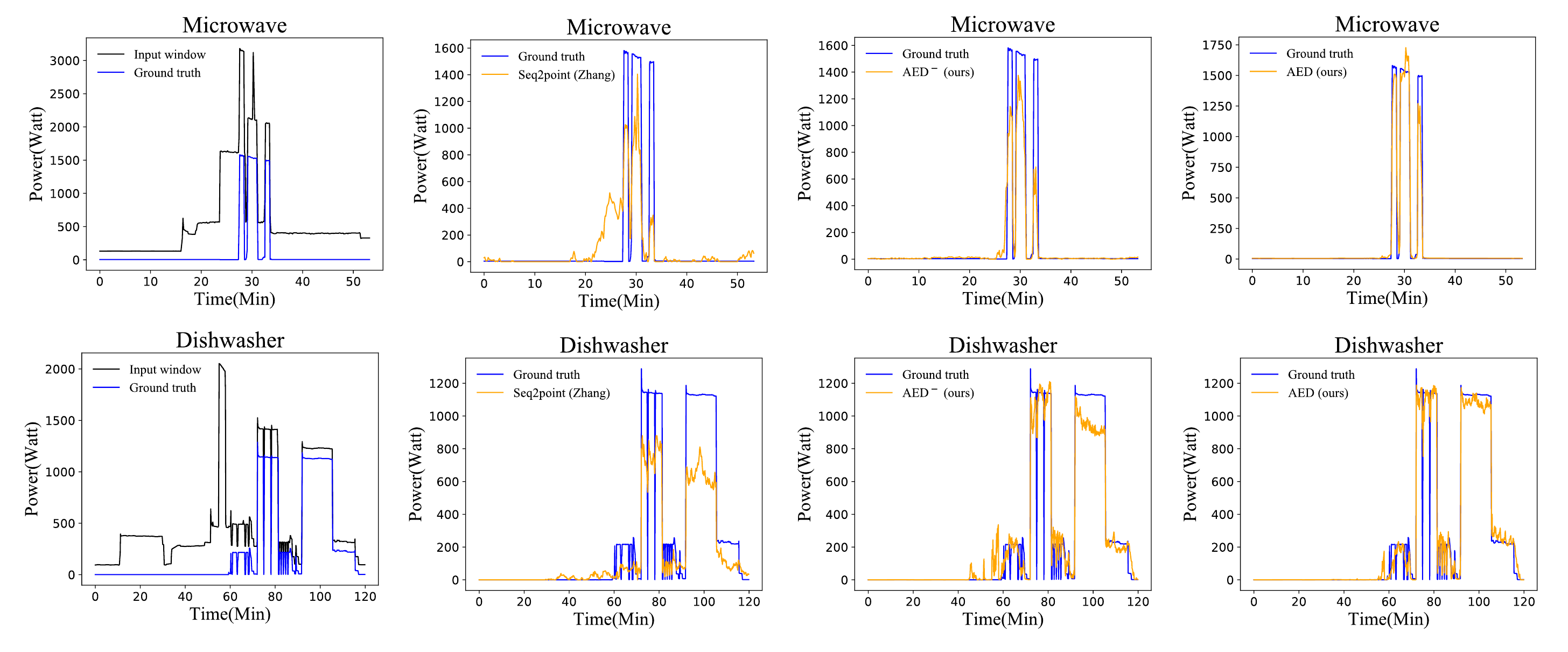}
  \end{center}
  \caption{Qualitative results on REDD. From left to right, the first column shows the results of mains readings and the ground truth of one specific appliance, e.g., dishwasher and microwave. The following columns report the results of seq2point, ours without multi-adversarial learning and our full model.}
  \label{fig:redd}
  \end{figure*}

\begin{figure*}[t]
  \begin{center}
  \includegraphics[width=0.99\linewidth]{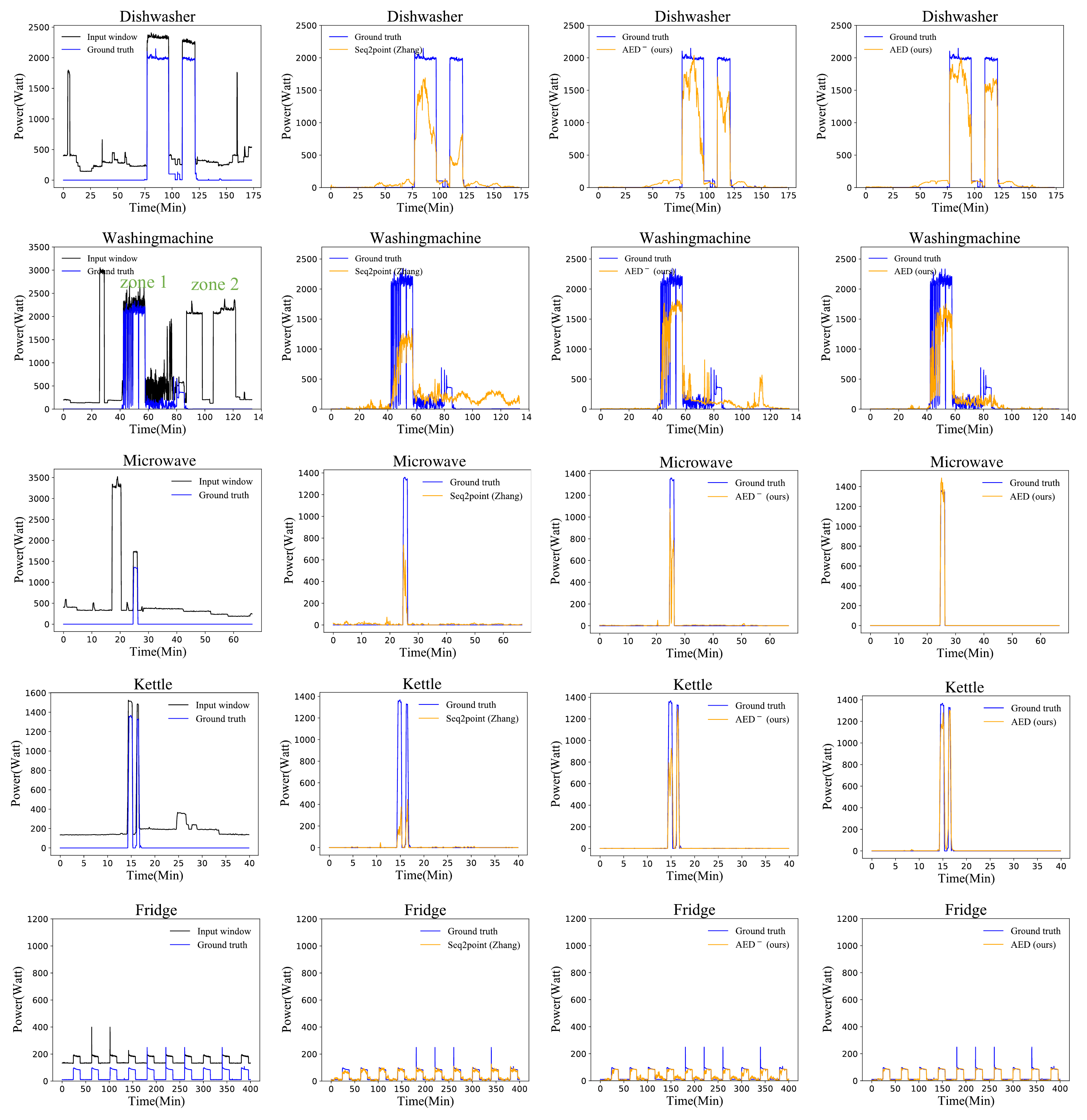}
  \end{center}
  \caption{Qualitative results on UK-DALE. From left to right, the first column shows the results of mains readings and the ground truth of one specific appliance. We choose all appliances in UK-DALE. The following columns report the results of seq2point, ours without multi-adversarial learning and our full model.}
  \label{fig:dale}
  \end{figure*}

\subsection{Qualitative Results}
For a better understanding of the proposed method, we further report the qualitative results of our method and seq2point~\cite{zhang2018sequence} in Fig.~\ref{fig:redd} and Fig.~\ref{fig:dale}. Specifically, we visualize the signals of different observations in a recording period. Notably, we report the results of our AED without multi-adversarial learning, which is denoted as AED$^-$ in the figures, in the third column, which can be seen as a result of ablation study. Comparing seq2point with our AED, it is easy to observe that our method can better fit the ground truth. From the results in the last columns and the third columns, we can see that the multi-adversarial learning can further improve the performance of AED. \textcolor{black}{Moreover, in the figures of fridge, the seq2point shows severe fluctuations, while the curve obtained by our method is considerably smooth, indicating that our model is very stable and is able to filter the noise introduced by other appliances effectively.}

In addition, we manually added two green boxes in the figures of washing machine in Fig.~\ref{fig:dale}. For the sake of narration, we name the areas in the left green box and right green box as zone~1 and zone~2, respectively. It can be seen from the mains readings that zone~1 and zone~2 are similar in a way, e.g., similar consumption value and similar duration. However, the consumption in zone~1 is contributed by the washingmachine (reflected by the ground truth line) and zone~2 is contributed by other appliances. It can be seen that our AED is able to distinguish such a subtle difference while other methods cannot, which verifies the superiority of our AED formulation.

\subsection{Model Discussion}
{\bf Ablation Study.} Our method consists of the feature representation part and the multi-adversarial learning part. Fig.~\ref{fig:redd} and Fig.~\ref{fig:dale} have reported the results of our method without the multi-adversarial learning, denoted as AED$^-$. In Table~\ref{tab:ablation}, we report the comprehensive ablation study of our model. It is obvious that multi-adversarial learning is beneficial to reduce MAE in all appliances and brings an averaged reduction on MAE by 1.76.

\begin{table*}
  \caption{Ablation study on REDD and UK-DALE datasets in terms of MAE. We take microwave and fridge of UK-DALE as examples.}
  \label{tab:ablation}
  \centering
  \begin{tabular}{llccccc}
    \toprule
    Dataset & Methods & Microwv & Fridge & Dishwsh & Washmch  & Average  \\
    \midrule
    \multirow{2}*{REDD}  & AED$^-$ & 21.253 & 31.122 & 13.896 & 12.791 & $19.76 \pm 7.32$\\
    \cmidrule(r){2-7}
                     & AED [Ours] & 17.914 & 29.770 & 11.527 &  11.533 & ${17.68 \pm 7.44}$\\
     \midrule
     \multirow{2}*{UK-DALE} & AED$^-$ & 6.924 & 15.546 & 17.642 & 6.985 & $11.78 \pm 6.09$\\
    \cmidrule(r){2-7}
                     & AED [Ours] & 5.948 & 13.324 & 15.062 & 5.764 & ${10.02 \pm 6.11}$  \\
     \bottomrule
  \end{tabular}
\end{table*}

\begin{figure*}[t]
  \begin{center}
  \subfigure[Training Process.]{
  \includegraphics[width=0.4\linewidth]{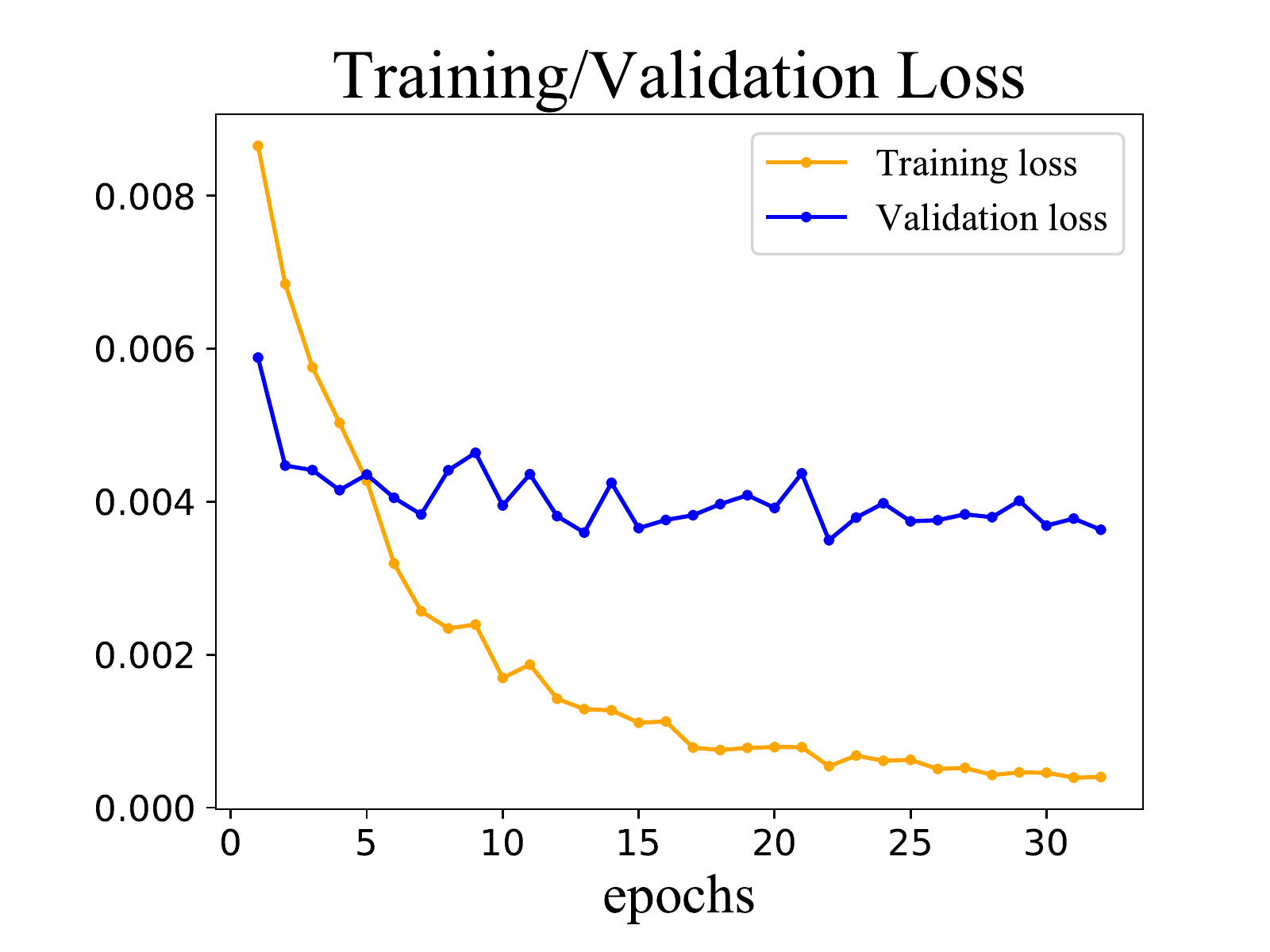}
  }
  \subfigure[Parameter Sensitivity.]{
  \includegraphics[width=0.4\linewidth]{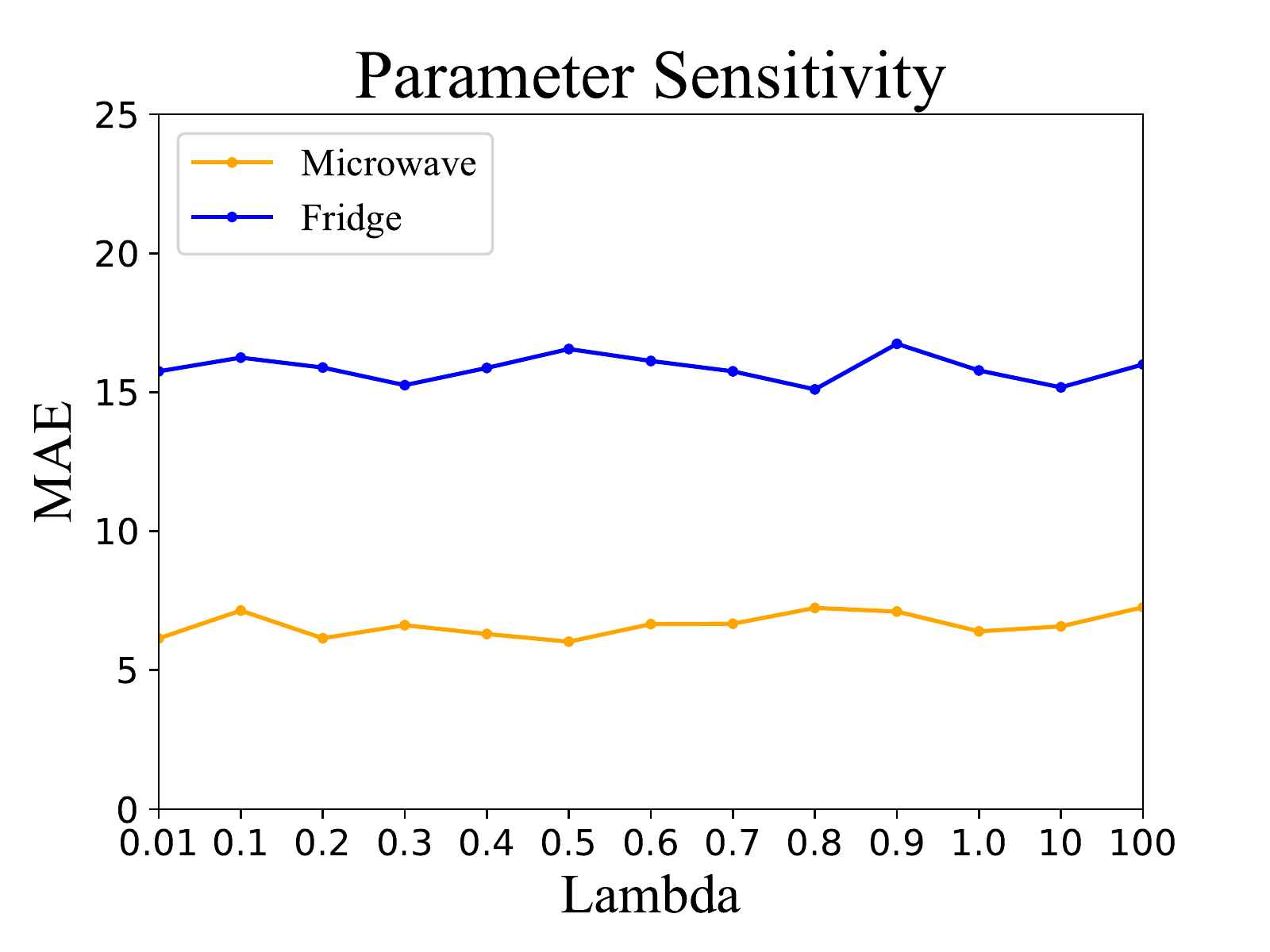}
  }
  \end{center}
  \caption{Model analysis. (a) Training process of the proposed method. We take the microwave of UK-DALE as an example. (b) Parameter sensitivity of the proposed method.}
  \label{fig:training}
  \end{figure*}

{\bf Training Process.} To show the training process of AED, we report both the validation loss and the training loss of AED on microwave of UK-DALE in Fig.~\ref{fig:training}(a). As an adversarial model, it can be seen that our method is relatively stable during the training and it can achieve convergence in a few epochs. 

{\bf Parameter Sensitivity.} Our method involves a hyper-parameter $\lambda$ which is tuned on the validation set. Fig.~\ref{fig:training}(b) reports the sensitivity of $\lambda$ on UK-DALE dataset. We turn the $\lambda$ from 0.01 to 100. It can be seen that $\lambda$ is not sensitive, our model has a high tolerance for the variation of hyper-parameter.



\section{Conclusion and Discussion}
In this paper, we propose a novel method named adversarial energy disaggregation (AED) for non-intrusive load monitoring (NILM). Specifically, we introduce the idea of adversarial learning into solving energy disaggregation problems. To learn better feature representations and capture the complex multimode structures of various appliances, we report a new CNN architecture and a multi-adversarial learning paradigm. Extensive experiments on two real-world datasets verify that our AED is able to achieve the new state-of-the-art results. In our proposed model, we need to pre-train several models, e.g., the discriminators, according to the number of appliances. Although it brings additional work, the total number of appliances is small for a typical family, and we can reuse the pre-trained models for different families. For instance, we can use transfer learning~\cite{li2020maximum,li2019locality,li2018heterogeneous} to challenge the problem. In our further work, we will explore to transfer the pre-trained model across different families and try to recognize unseen appliances.

At last, the proposed method in this paper is a non-intrusive monitoring method, which means one can predict the electric appliances of a family without access into the house. For some reasons, the electric appliances in a family could be private information. On the other hand, patterns of energy use can reflect the behavior patterns of human beings. For instance, the states of lights, e.g., turn on or off, could indicate whether there is people at home. Thus, it is suggested to use the technology under the condition of being fully acknowledged by the resident.

\section{Acknowledgement}
This work was supported in part by the National Natural Science Foundation of China under Grant 61806039 and 62073059, and in part by Sichuan Science and Technology Program under Grant 2020YFG0080 and 2020YFG0481.

\bibliographystyle{ACM-Reference-Format}
\bibliography{neurips_2020}


\end{document}